\documentclass[aps,prd,superscriptaddress,twoside,twocolumn,nofootinbib,%
10pt,showpacs,floatfix]{revtex4-1}

\usepackage{amsmath,amssymb}
\usepackage{graphicx,bm}
\usepackage{slashed}
\usepackage{dcolumn}
\usepackage{multirow}

\allowdisplaybreaks

\begin{document}


\title{Hexaquark picture for $d^*(2380)$}


\author{Hungchong Kim}%
\email{hungchong@kau.ac.kr}
\affiliation{Research Institute of Basic Science, Korea Aerospace University, Goyang, 412-791, Korea}
\affiliation{Center for Extreme Nuclear Matters, Korea University, Seoul 02841, Korea}

\author{K. S. Kim}%
\affiliation{School of Liberal Arts and Science, Korea Aerospace University, Goyang, 412-791, Korea}

\author{Makoto Oka}%
\affiliation{Advanced Science Research Center, Japan Atomic Energy Agency, Tokai, Ibaraki, 319-1195, Japan}
\affiliation{Nishina Center for Accelerator-Based Science, RIKEN, Wako 351-0198, Japan}

\date{\today}


\begin{abstract}

Hexaquark wave function with the quantum numbers $I(J^P)=0(3^+)$, which might be relevant for $d^*(2380)$,
is constructed under an assumption that this is composed only of $u,d$ quarks in an $S$-wave.
By combining three diquarks of either type, ($\bm{\bar{3}}_c, I=1$) or ($\bm{6}_c, I=0$),
we demonstrate that there are five possible configurations for the six-quark state.
The fully antisymmetric wave function is constructed by linearly combining the five configurations on an equal footing.
We then take this wave function as well as the five configurations
to calculate the hexaquark mass using
the contact type effective potential consisting of the color-spin, color electric and constant shift.
The mass is found to be the same regardless of the configurations being used including the fully antisymmetric one.
This result can be traced to the fact that the hexaquark system has a freedom in choosing
three diquarks in the construction of its wave function.
The calculated hexaquark mass using the empirical parameters independently fixed from the baryon spectroscopy
is found to be around $2342$ MeV, which is indeed very close to the experimental mass of $d^*(2380)$.
Therefore, the hexaquark picture is promising for $d^*(2380)$ as far as the mass is concerned.

\end{abstract}

\maketitle

\section{Introduction}

The $d^*(2380)$ resonance with the quantum numbers $I(J^P)=0(3^+)$ has been reported recently by
WASA-at-COSY Collaboration~\cite{Adlarson:2011bh} from the exclusive reaction channel,
$pn\rightarrow d\pi^0\pi^0$.
Its existence is also supported by later experiments~\cite{Adlarson:2012au,Adlarson:2013usl,
Adlarson:2014pxj,Adlarson:2014xmp,Adlarson:2014tcn}.
This resonance, as a state of double-pionic fusion to deuterium,
may provide a plausible explanation
for the substantial enhancement seen long ago in the $^3\text{He}$ missing mass spectrum from the
inclusive reaction, $pd\rightarrow ^3$$\text{He} X$~\cite{ABC:1960,ABC:1961}.

Based on the reaction channels that lead to its observation, $d^*(2380)$ is expected to be a six-quark state
composed of $u,d$ quarks only. Then, as often encountered in the multiquark studies, one important issue to
discuss is whether $d^*(2380)$ is a molecular state of two color-singlet objects, namely the anticipated dibaryon state, or
a hexaquark state that has the hidden-color component in addition.
The $d^*(2380)$ mass, which is measured to be around $M\approx 2380$ MeV,
is about 80 MeV less than the invariant mass of $\Delta\Delta$.
In this sense, $d^*(2380)$ could be the dibaryon state
predicted long ago by Ref.~\cite{Dyson:1964xwa} with the mass 2350 MeV in the isospin-spin channel of $(I,J)=(0,3)$.
The dynamics that leads to this resonance could be the attractive force
between two $\Delta$s in the channel $(I,J)=(0,3)$~\cite{Oka:1980ax}.

One problem in this view is the small decay width of $d^*(2380)$ which is about 70 MeV~\cite{Adlarson:2011bh}.
Given the fact that the $\Delta$ decay width, $\Gamma_\Delta\approx 115$ MeV,
it is not easy to understand the small width if $d^*(2380)$ is viewed as a bound state of two $\Delta$s.
Alternatively, as advocated in Refs.~\cite{Huang:2014kja, Huang:2015nja},
$d^*(2380)$ might be a hexaquark state that is dominated by the hidden-color component.
Indeed, according to Refs.~\cite{Bashkanov:2013cla, Harvey:1988nk}, the six-quark state with all the quarks
in an $S$-wave is found to have more probability to stay
in the hidden-color configuration rather than in the $\Delta\Delta$ configuration.
In this regard, the hexaquark picture needs to be investigated more concretely because, after all, it is more extensive
in that this picture can accommodate the molecular picture as its component.

To investigate the hexaquark possibility, Ref.~\cite{Park:2015nha} has performed
a calculation based on a variation method and concluded that the hexaquark picture
is not realistic for $d^*(2380)$ because the calculated mass from this picture is too large.
This conclusion is questioned later by Ref.~\cite{Lu:2017uey}
which claims that the important medium-range interaction was missing in that calculation.
More recently, Shi et al.~\cite{Shi:2019dpo} investigated the hexaquark possibility for $d^*(2380)$ in a model
where the spin-1 diquark with the color structure of $\bm{\bar{3}}_c$ has been adopted
in the construction of the hexaquark wave function. Even though this spin-1 diquark is known to be the bad diquark~\cite{Jaffe04},
their calculation seems to reproduce the experimental mass and decay width of $d^*(2380)$ relatively well.
This result, however, was disputed later by Ref.~\cite{Gal:2019xju} which pointed out some
problems in calculating the decay width and an unrealistic nature inherited in the calculated mass.

The diquark approach, which normally relies on a compact diquark with a hope to generate an optimal multiquark configuration,
has conceptual problems especially when it applies to the hexaquark system of $qqqqqq$, ($q=u,d$).
First of all, such a hexaquark, as it is built from one diquark type only,
cannot satisfy the fully antisymmetric condition under exchange of any two-quark among the six quarks~\cite{Park:2015nha,Gal:2019xju}.
Secondly, the resulting hexaquark configuration does not have a privilege over
other possible configurations constructed from other diquark types as far as
in generating the ground state configuration.
The second statement is closely related to a freedom in dividing the six quarks into three diquarks.
There are various ways to divide the six quarks into three diquarks~\footnote{This is in
contrast to the tetraquark system $qq\bar{q}\bar{q}$ which has only one division $qq$-$\bar{q}\bar{q}$ in a diquark model.
The other like $q\bar{q}$-$q\bar{q}$ is not a division based on a diquark model.}
and, in principle, all of them must be equivalent to describe the hexaquark system
as it is completely arbitrary to choose any division especially in the $qqqqqq$ system with $q=u,d$ only.
It turns out that, if all the possible diquarks are considered in the hexaquark construction,
there are five configurations in each division and, from the freedom mentioned above, it is possible to show
that all the five configurations have the same expectation value for the potential (see Sec.~\ref{sec:veff}).
As a result, all the five configurations are equally important and none of them is in fact better suited for the hexaquark description.
In other words, there is no compact diquark that can generate the compact hexaquark configuration.
Instead, the five configurations can participate in constructing
the fully antisymmetric wave function as its components.
Therefore, even though the diquark approach is conceptually problematic, it still provides
one convenient basis to construct a physical wave function for the hexaquark.

In this work, we investigate the hexaquark possibility for $d^*(2380)$.
The hexaquark wave functions will be constructed under an assumption that all the quarks are in an $S$-wave.
In our construction, we take all the possible diquarks as a convenient tool for describing
the hexaquark wave function so the diquark types are not necessarily limited to the compact
one as in the usual diquark models.
In this sense, our approach is different from other studies that rely on one compact diquark type only.
In fact, there are five possible configurations that can be combined to form a fully antisymmetric wave function.
The resulting hexaquark wave function will be tested by calculating its mass
using a semiempirical effective potential of the contact type composed of color-spin, color-electric and some constant shift,
the same potential type that has been used in the previous studies of tetraquarks~\cite{Kim:2016dfq,Kim:2017yur,Kim:2017yvd,Kim:2018zob,Lee:2019bwi}.
The parameters appearing in this potential will be independently fixed from the baryon octet and decuplet.
Thus, our approach in this work is different from Ref.~\cite{Park:2015nha} in that
we use this contact type potential
which freezes the spatial dependence of the potential by fitting the effective parameters from the baryon spectroscopy
handled in the same footing.

This paper is organized as follows.
In Sec.~\ref{sec:potential}, we introduce an effective potential that will be used in our study.
In Sec.~\ref{sec:wave}, all the possible hexaquark configurations are constructed in color,
isospin space separately and also in the combined color-isospin space.
We also present explicit expression for the fully antisymmetric wave function
in Sec.~\ref{subsec:full}. In Sec.~\ref{sec:veff}, we make a few remarks on the interesting aspect from the expectation value of the effective potential
when it is calculated with respect to the constructed hexaquark wave functions. The hexaquark mass will be presented in Sec.~\ref{sec:mass}.
We summarize in Sec.~\ref{sec:summary}.

\section{Effective potential}
\label{sec:potential}

We begin with an effective potential that will be used to calculate the hexaquark mass.
A hadron in the constituent quark picture can be described by the Hamiltonian composed of two terms,
the quark mass term and the interaction term among the participating quarks.
The interaction can have two different sources, one-gluon exchange potential and
instanton-induced potential~\cite{OT89,Oka:1990vx}.
One way to parameterize the effective potentials is
to write them down in the contact form composed of the three parts, color-spin($V_{CS}$), color-electric($V_{CE}$),
and constant shift~\cite{Kim:2016tys},
\begin{eqnarray}
V_{eff}&& = V_{CS}+V_{CE}+\text{constant}\nonumber \\
=&&\sum_{i < j} \frac{v_0}{m_i m_j} \lambda_i \cdot \lambda_j J_i\cdot J_j\!
+\!\sum_{i < j} \frac{v_1}{m_i m_j} \lambda_i \cdot \lambda_j\! + v_2 .
\label{interaction}
\end{eqnarray}
Here $\lambda_i$ denotes the Gell-Mann matrix for the color, $J_i$ the spin, $m_i$ the quark mass.
This interaction is a semiempirical type which acts on two quarks in one spatial point.
So the spatial dependence of $V_{eff}$ is frozen in an average sense by fixing the empirical parameters, $v_0,v_1,v_2$,
from some baryon masses used as inputs.

Hadron mass can be written formally by the mass formula
\begin{equation}
M_H \simeq \sum_{i} m_i^{} + \langle V_{eff} \rangle\ ,
\label{mass}
\end{equation}
where the expectation value needs to be evaluated with respect to an appropriate wave function constructed for the hadron of concern.
In the constituent quark picture, the quark mass used in Eq.~(\ref{mass})
should be regarded as an effective mass that includes the kinetic energy of constituent quarks~\cite{Shi:2019dpo}.

The color-spin interaction $V_{CS}$ alone is often used to investigate
hadron masses~\cite{Zeldovich:1967rt, Karliner:2003dt, Karliner:2006fr, Ali:2019roi, Maiani:2004vq} in the context of
Eq.~(\ref{mass}).
One advantage of using $V_{CS}$ is that it
reproduces quite well the mass difference among hadrons with the same
flavor content~\cite{Kim:2016dfq,Kim:2017yvd,DeRujula:1975qlm,Keren07,Silve92,GR81,Lee:2009rt}
because the other terms in the potential, the color-electric and constant shift, are canceled away in the mass difference.
But in general the other terms are not negligible in calculating the mass itself.
Indeed, the mass formula, Eq.~(\ref{mass}), with all the three terms kept in the potential, has been applied
to the baryon system successfully~\cite{Kim:2016tys} using the
three parameters, $v_0,v_1,v_2$, fitted from the experimental masses of $N,\Delta, \Lambda$.
To make our presentation self-contained, this fitting process has been explained in the Appendix~\ref{sec:baryon}
with the three different cases, I) $v_0\ne 0, v_1=v_2=0$, II) $v_0\ne 0, v_1\ne 0, v_2=0$, III) $v_0\ne 0, v_1\ne 0, v_2\ne 0$.
We find that the third case with the determined parameters
\begin{eqnarray}
v_0&=&(-199.6~\text{MeV})^3,\  v_1=(71.2~\text{MeV})^3\nonumber\ ,\\
v_2&=&122.5~\text{MeV}\ ,
\label{parameters}
\end{eqnarray}
reproduces the baryon masses very well as shown in Table~\ref{baryon masses} in Appendix~\ref{sec:baryon}.

The effective potential, Eq.~(\ref{interaction}), can be simplified further
when it applies to the hexaquark system of our concern.
As mentioned already, we consider in this work the hexaquark composed of $u,d$ quarks only
so we can set all the quark masses to be equal, $m_i=m_j\equiv m$.
Its value is taken to be 330 MeV as in the previous works on tetraquarks~\cite{Kim:2016dfq,Kim:2017yur,Kim:2017yvd,Kim:2018zob}.
Also, all the six quarks are in the spin-up state in order to make the total spin $J=3$.
Hence, the spin-dependent part in
Eq.~(\ref{interaction}) is trivially evaluated to be $\langle J_i\cdot J_j\rangle=1/4$ for any $i,j$.
These two aspects simplify Eq.~(\ref{interaction}) further into the from
\begin{eqnarray}
V_{eff} = \left [ \frac{v_0}{4m^2} + \frac{v_1}{m^2} \right ]\sum_{i < j} \lambda_i \cdot \lambda_j  + v_2\ .
\label{eff}
\end{eqnarray}
Now only nontrivial part in evaluating $\langle V_{eff} \rangle$ is the color-color part
$\sum_{i < j} \lambda_i \cdot \lambda_j$ with respect to an appropriate hexaquark
wave function that will be constructed in the next section.
Since this effective potential does not depend on isospin, $\langle V_{eff} \rangle$
is practically independent of various isospin configurations that lead to the total isospin, $I=0$.

\section{Hexaquark wave functions}
\label{sec:wave}

In this section, we construct a hexaquark wave function with the quantum numbers, $I(J^P)=0(3^+)$.
Since all the six quarks are assumed to be in an $S$-wave, the spatial part
is fully symmetric under exchange of any two quarks.  The spin part is also symmetric because
all the quarks are in the spin-up state.  Then, the rest color-isospin part must be antisymmetric.
To achieve this, we start with all the possible diquark types that obey the Pauli principle within the two quarks
and use them as a convenient tool to construct
the five possible configurations for $qq-qq-qq$ in the color-isospin part.
The resulting configurations therefore are antisymmetric only between the two quarks in each diquark.
The fully antisymmetric wave function will be constructed by linearly combining all the five configurations.

\subsection{Color part}
\label{subsec:color}

We start with the color part of the hexaquark wave function by combining three diquarks of all the possible types.
For an illustrative purpose, we label the six quarks by $q_1 q_2 q_3 q_4 q_5 q_6$ and divide them
into three diquarks grouped as (12)(34)(56).
The six-quark colors
can be expressed by the diquark colors in this division of (12)(34)(56) as
\begin{eqnarray}
&&\left\{[\bm{3}_c\otimes \bm{3}_c] \otimes [\bm{3}_c \otimes \bm{3}_c] \otimes[\bm{3}_c \otimes\bm{3}_c]\right\}_{\bm{1}_c}\nonumber\ \\
&&=\left\{ [\bm{6}_c \oplus \bm{\bar{3}}_c]\otimes [\bm{6}_c \oplus \bm{\bar{3}}_c]
\otimes [\bm{6}_c \oplus \bm{\bar{3}}_c]\right\}_{\bm{1}_c}\label{6q}\ ,
\end{eqnarray}
using the group multiplication of $\bm{3}_c \otimes \bm{3}_c = \bm{6}_c \oplus \bm{\bar{3}}_c$.
Here the subscript $\bm{1}_c$ of the total bracket denotes that the hexaquark is in a color singlet.

It is now easy to see that, among the various terms that Eq.~(\ref{6q}) can generate,
only {\it five} color configurations can form a color-singlet state totally.
This coincides with the general statement that a six-quark state has five color configurations no matter how it is divided
into subparts~\cite{Brodsky:1983vf,Ji:1985ky}.
The five color configurations in this diquark division can be written explicitly as
\begin{eqnarray}
&&\left\{[\bm{6}_c\otimes \bm{6}_c]_{\bar{\bm{6}}_c}\otimes \bm{6}_c\right\}_{\bm{1}_c}\label{color1},\\
&&\left\{[\bm{\bar{3}}_c\otimes \bm{\bar{3}}_c]_{\bm{3}_c}\otimes \bm{\bar{3}}_c\right\}_{\bm{1}_c}\label{color2},\\
&&\left\{[\bm{6}_c\otimes \bm{\bar{3}}_c]_{\bm{3}_c}\otimes \bm{\bar{3}}_c\right\}_{\bm{1}_c}\label{color3},\\
&&\left\{[\bm{\bar{3}}_c\otimes\bm{6}_c]_{\bm{3}_c}\otimes \bm{\bar{3}}_c\right\}_{\bm{1}_c}\label{color4},\\
&&\left\{[\bm{\bar{3}}_c\otimes \bm{\bar{3}}_c]_{\bar{\bm{6}}_c}\otimes \bm{6}_c\right\}_{\bm{1}_c}\label{color5} .
\end{eqnarray}
To explain our notation, $[\bm{6}_c\otimes \bm{6}_c]_{\bar{\bm{6}}_c}$ in Eq.~(\ref{color1}) denotes the four-quark state
belonging to ${\bar{\bm{6}}_c}$ constructed from the diquark $(12) \in \bm{6}_c$ and the other diquark $(34) \in \bm{6}_c$.
This four-quark state is then combined with the third diquark $(56) \in \bm{6}_c$ to form a color-singlet state totally.

The five color configurations above have been expressed by the two possible diquark types in color, the symmetric one $\bm{6}_c$
and the antisymmetric one $\bm{\bar{3}}_c$.
Both diquarks are of course in the spin-1 state by construction.
Note that the diquark ($\bm{6}_c, J=1$) is lower in potential (i.e. more stable) than the ($\bm{\bar{3}}_c, J=1)$ diquark
if the binding is calculated from $V_{eff}$ using the parameter set given in
Eq.~(\ref{parameters}).
Thus, the hexaquark configuration of Eq.~(\ref{color1}) is constructed from the stable diquark $\bm{6}_c$ only
while Eq.~(\ref{color2}), which has been used in Ref.~\cite{Shi:2019dpo}, is built from the less stable diquark $\bm{\bar{3}}_c$ only.
The rest three configurations, Eqs.~(\ref{color3}),(\ref{color4}),(\ref{color5}) contain both types, $\bm{6}_c$ and $\bm{\bar{3}}_c$.

For a mathematical convenience, it is useful to represent
the five color configurations Eqs.~(\ref{color1})$\cdot\cdot\cdot$~(\ref{color5}) by a tensor notation.
In this notation\footnote{For technical details on this tensor notation, see Ref.~\cite{Oh:2004gz}},
$\bm{6}_c$ and $\bm{\bar{3}}_c$ are expressed by individual quark color as
\begin{eqnarray}
\bm{6}_c:~ S_{ab} &=& \frac{1}{\sqrt{2}}\left [ q_{a} q_{b} + q_{b} q_{a} \right ]\ ,\\
\bm{\bar{3}}_c:~ T^{a} &=& \frac{1}{\sqrt{2}}\epsilon^{abc}\left [ q_{b} q_{c} - q_{c} q_{b} \right ]=\sqrt{2}\epsilon^{abc}q_bq_c\ .
\end{eqnarray}
From this expression, one can explicitly see that $S_{ab}$ is symmetric and $T^{a}$ is antisymmetric
under exchange of the quark colors.
Their inner products are normalized as
\begin{eqnarray}
(S_{ab},S_{cd}) &=& \delta_{ac}\delta_{bd}+\delta_{ad}\delta_{bc}\label{6tensor}\ ,\\
(T^a,T^b) &=& 4\delta^{ab}\label{3tensor}\ ,
\end{eqnarray}
and they are orthogonal,
\begin{eqnarray}
(S_{ab}, T^c)=0\ .
\end{eqnarray}

Now, it is straightforward to write Eqs.~(\ref{color1})$\cdot\cdot\cdot$(\ref{color5}) in terms of $S_{ab},T_a$, namely,
\begin{eqnarray}
|\text{C}_1\rangle &=& \frac{1}{12}\epsilon^{abc}\epsilon^{a'b'c'}(S_{12})_{aa'}(S_{34})_{bb'}(S_{56})_{cc'}\label{c1},\\
|\text{C}_2\rangle &=& \frac{1}{8\sqrt{6}}\epsilon_{abc} (T_{12})^a (T_{34})^b (T_{56})^c\label{c2}, \\
|\text{C}_3\rangle &=& \frac{1}{8\sqrt{3}} (S_{12})_{ab} (T_{34})^a (T_{56})^b\label{c3}\ ,\\
|\text{C}_4\rangle &=& \frac{1}{8\sqrt{3}} (T_{12})^a (S_{34})_{ab} (T_{56})^b\label{c4}\ ,\\
|\text{C}_5\rangle &=& \frac{1}{8\sqrt{3}} (T_{12})^a (T_{34})^{b} (S_{56})_{ab}\label{c5}\ .
\end{eqnarray}
In the right hand side (RHS), we have added the numerical subscripts in our tensors in order to specify the division (12)(34)(56) more
clearly~\footnote{Our expression for $|\text{C}_i\rangle$
is different from Eq.(21) in Ref.~\cite{Park:2015nha} in that ours are written in a diquark basis.}.
These five color configurations are orthonormal,
$\langle \text{C}_i|\text{C}_j\rangle =\delta_{ij}$ .

The division (12)(34)(56) above is just one particular choice in representing the color configurations
of the hexaquark.
Since how dividing the six quarks into three diquarks is completely arbitrary,
one can choose a different division like (13)(24)(56) that can be obtained from (12)(34)(56) by $q_2 \leftrightarrow q_3$.
The equation like Eq.~(\ref{6q}) still holds in this new division and one can get similar five configurations
that differ from $|\text{C}_i\rangle$ only by the numerical subscripts in the RHS.
But it is clear that $|\text{C}_i\rangle$ should be related to these new configurations
through some linear combinations or vice versa
because any diquark that is in a definite color state in the (12)(34)(56)
division is in a mixture of $\bm{6}_c$ and $\bm{\bar{3}}_c$ when it is viewed in the different division like (13)(24)(56).
This aspect can be utilized to prove that there are no lowest configuration among the five.
We will come back to this discussion later when we calculate the expectation value, $\langle V_{eff}\rangle$, in Sec.~\ref{sec:veff}.

\subsection{Isospin part}
\label{subsec:isospin}

Next we develop the isospin part of the wave function that needs to be combined into the
color part, $|\text{C}_i\rangle$, in Eqs.~(\ref{c1})$\cdot\cdot\cdot$~(\ref{c5}).
The isospin part is crucial for constructing the fully antisymmetric wave function
under exchange of any two-quark among the six quarks but,
since the effective potential $V_{eff}$ is blind on isospin, the isospin part does not practically participate
in calculating $\langle V_{eff}\rangle$.

In our construction, we first impose the antisymmetric constraint only on each diquark
in the combined space of color-isospin. Any diquark in the configurations $|\text{C}_i\rangle$
is either symmetric ($S_{ab}$) or antisymmetric ($T^c$) in color.
To make each diquark antisymmetric in the color-isospin space, the diquark isospin ($I_{di}$) is restricted to be
$I_{di}=0$ for the $S_{ab}$ diquark and $I_{di}=1$ for the $T^c$ diquark.
Then the total isospins are determined
from the multiplication of three isospins of the diquarks.
Specifically, $|\text{C}_1\rangle$ in Eq.~(\ref{c1}) consists of three diquarks of the $S_{ab}$ type with the isospin $I_{di}=0$.
Consequently, the total isospin of $|\text{C}_1\rangle$ is $I=0$ only.
For $|\text{C}_2\rangle$ of Eq.~(\ref{c2}), each diquark is in the $I_{di}=1$ state
so possible isospins of $|\text{C}_2\rangle$ are $I=0,1,2,3$.
For $|\text{C}_3\rangle, |\text{C}_4\rangle, |\text{C}_5\rangle$, possible isospins are $I=0,1,2$.
This means, the $I=0$ is the only common isospin state that exists in all the five configurations.
As we shall see below, since all the five color configurations are necessary in constructing the fully antisymmetric wave function,
the $I=0$ is the only possible isospin for the hexaquark and this is indeed consistent with the isospin of $d^*(2380)$.
This observation here provides an alternative explanation why there is only one isospin state, $I=0$,
when the six-quark state is in the spin state $J=3$ totally~\cite{Dyson:1964xwa}.

It is straightforward to derive the five isospin configurations, with the total isospin $I=0$,
that can be multiplied to the corresponding five color configurations,
Eqs.~(\ref{c1})$\cdot\cdot\cdot$~(\ref{c5}) respectively. They are
\begin{eqnarray}
|I_1\rangle &=& [ud][ud][ud]\label{I1}\ ,\\
|I_2\rangle &=& \frac{1}{\sqrt{6}}\Big [ \big(uu\{ud\} - \{ud\}uu\big)dd - \big(uudd-dduu\big)\{ud\} \nonumber \\
 && + \big(\{ud\}dd-dd\{ud\}\big) uu \Big ]\label{I2}\ ,\\
|I_3\rangle &=& \frac{1}{\sqrt{3}}\Big ([ud] uudd - [ud]\{ud\}\{ud\} + [ud]dduu\Big )\label{I3}\ ,\\
|I_4\rangle &=& \frac{1}{\sqrt{3}}\Big ( uu[ud]dd - \{ud\}[ud]\{ud\} + dd[ud]uu\Big )\label{I4}\ ,\\
|I_5\rangle &=& \frac{1}{\sqrt{3}}\Big ( uudd[ud] - \{ud\}\{ud\}[ud] + dduu[ud]\Big )\label{I5}\ ,
\end{eqnarray}
where we have introduced the short-hand notations,
\begin{eqnarray}
[ud] \equiv \frac{1}{\sqrt{2}}(ud-du) ;\
\{ud\} \equiv \frac{1}{\sqrt{2}}(ud+du)\ ,
\end{eqnarray}
to represent the antisymmetric ($I_{di}=0$) and symmetric ($I_{di}=1$) combination respectively.
These five isospin configurations are also orthonormal,
$\langle I_i|I_j\rangle =\delta_{ij}$ .

\subsection{Color-isospin part}
\label{subsec:color-isospin}

The color-isospin part of the hexaquark wave function can be constructed from the direct product,
\begin{eqnarray}
|\psi_i\rangle = |\text{C}_i \rangle \otimes |I_i \rangle ,~~
(i=1,2,3,4,5)\label{CI}\ .
\end{eqnarray}
As we have mentioned already, to make each diquark antisymmetric in the combined space of color-isospin,
the product here must acts on only between a diquark in color and the corresponding diquark in isospin,
that is, the (12) diquark in color must be combined only with the (12) diquark in isospin and so on.
To show what we meant explicitly, $|\psi_5\rangle$, which is obtained by multiplying Eq.~(\ref{c5}) and Eq.~(\ref{I5}), takes the form
\begin{eqnarray}
|\psi_5\rangle &\propto&  (T_{uu})_a (T_{dd})_b (S_{[ud]})^{ab} - (T_{\{ud\}})_a (T_{\{ud\}})_b (S_{[ud]})^{ab}\nonumber \\
&&+ (T_{dd})_a (T_{uu})_b (S_{[ud]})^{ab} \label{CI5}\ .
\end{eqnarray}
Here $(T_{uu})_a$, $(S_{[ud]})^{ab}$ in the first term are defined as
\begin{eqnarray}
&&(T_{uu})_a=(T_{12})_{a}\otimes uu=\sqrt{2}\epsilon_{abc}u^b u^c\ ,\\
&&(S_{[ud]})_{aa'} = (S_{56})_{aa'}\otimes [ud]\nonumber \\
&&~~~= \frac{1}{2}(u_{a} d_{a'}+u_{a'}d_{a})-\frac{1}{2}(d_{a} u_{a'}+d_{a'}u_{a})\ ,
\end{eqnarray}
and the other terms are similarly defined.
Again, the five color-isospin wave functions are orthonormal,
$\langle \psi_i|\psi_j\rangle =\delta_{ij}$.

\subsection{Fully antisymmetric color-isospin part}
\label{subsec:full}

By construction, all the five color-isospin configurations $|\psi_i\rangle$ in Eq.~(\ref{CI})
is antisymmetric only under exchange of the two quarks in each diquark.
None of $|\psi_i\rangle$ is fully antisymmetric under exchange of any two-quark among all the six quarks and,
therefore, none of $|\psi_i\rangle$ can be regarded as the physical state.
But they constitute the full components of the hexaquark wave function because
one can construct the fully antisymmetric wave function $|\Psi\rangle$ by linearly combining all the five configurations as
\begin{eqnarray}
|\Psi\rangle = \sum_{i=1}^{5} a_i |\psi_i\rangle\ .
\end{eqnarray}
We determine the coefficients $a_i$ by two conditions, the normalization of the full wave function
and the antisymmetric constraint imposed on any two quarks in the six quarks.
This has been worked out explicitly and we find that
\begin{equation}
a_1=a_2=-a_3=-a_4=-a_5=\frac{1}{\sqrt{5}}\ .
\end{equation}
Therefore, the fully antisymmetric wave function in the color-isospin space is given as
\begin{eqnarray}
|\Psi\rangle = \frac{1}{\sqrt{5}}\left[ |\psi_1\rangle+|\psi_2\rangle-|\psi_3\rangle-|\psi_4\rangle-|\psi_5\rangle \right ]\label{full}\ .
\end{eqnarray}
From this expression, we see that
all the five color-isospin configurations $|\psi_i\rangle$ are equally important in making the final $|\Psi\rangle$ fully antisymmetric.

Finally, before closing this subsection, we present the explicit expression of this fully antisymmetric wave function given as
\begin{widetext}
\begin{eqnarray}
|\Psi\rangle &=& \frac{1}{12\sqrt{5}}\epsilon_{abc}\epsilon_{a'b'c'} (S_{[ud]})^{aa'}(S_{[ud]})^{bb'}(S_{[ud]})^{cc'} \nonumber\\
             &+&\frac{1}{48\sqrt{5}}\epsilon_{abc}\Big[ (T_{uu})^a (T_{\{ud\}})^b (T_{dd})^c  -(T_{\{ud\}})^a (T_{uu})^b (T_{dd})^c
             -(T_{uu})^a (T_{dd})^b (T_{\{ud\}})^c \nonumber\\
             &&~~+(T_{dd})^a (T_{uu})^b (T_{\{ud\}})^c +(T_{\{ud\}})^a (T_{dd})^b (T_{uu})^c -(T_{dd})^a (T_{\{ud\}})^b (T_{uu})^c\Big]\nonumber \\
             &-&\frac{1}{24\sqrt{5}}\Big[ (S_{[ud]})^{ab} (T_{uu})_a (T_{dd})_b - (S_{[ud]})^{ab} (T_{\{ud\}})_a (T_{\{ud\}})_b+(S_{[ud]})^{ab} (T_{dd})_a (T_{uu})_b\Big]\nonumber \\
             &-&\frac{1}{24\sqrt{5}}\Big[ (T_{uu})_a (S_{[ud]})^{ab} (T_{dd})_b - (T_{\{ud\}})_a (S_{[ud]})^{ab} (T_{\{ud\}})_b + (T_{dd})_a (S_{[ud]})^{ab}  (T_{uu})_b\Big]\nonumber \\
             &-&\frac{1}{24\sqrt{5}}\Big[ (T_{uu})_a (T_{dd})_b (S_{[ud]})^{ab} - (T_{\{ud\}})_a (T_{\{ud\}})_b (S_{[ud]})^{ab}  + (T_{dd})_a (T_{uu})_b (S_{[ud]})^{ab}  \Big]\label{explicit}\ .
\end{eqnarray}
\end{widetext}

\section{A few remarks on $\langle V_{eff} \rangle$}
\label{sec:veff}

To compute the hexaquark mass through Eq.~(\ref{mass}), first we need to calculate
the expectation value of $V_{eff}$ [Eq.~(\ref{eff})] with respect to the fully antisymmetric wave function, $\langle \Psi| V_{eff}|\Psi\rangle$.
Since $V_{eff}$ is blind on isospin, the orthonormal condition $\langle I_i|I_j\rangle =\delta_{ij}$ guarantees
that
\begin{eqnarray}
\langle \psi_i|V_{eff}|\psi_j\rangle =0\ ,~~(i\ne j)\label{no mixing}
\end{eqnarray}
and, for $i=j$,
\begin{eqnarray}
\langle \psi_i| V_{eff}|\psi_i\rangle=\langle \text{C}_i | V_{eff}|\text{C}_i\rangle &\equiv V_i\ .
\label{effp}
\end{eqnarray}
The last equation also defines $V_i$, the expectation values of $V_{eff}$ with respect to $|\psi_i\rangle$.

One important characteristics on $\langle V_{eff} \rangle$ is that
\begin{eqnarray}
\langle \Psi| V_{eff}|\Psi\rangle = V_1=V_2=\cdot\cdot\cdot=V_5\ .
\label{Equiv}
\end{eqnarray}
This basically says that $|\Psi\rangle$ as well as $|\psi_i\rangle$ have the same expectation value of the potential.
This also shows that the fully antisymmetric state $|\Psi\rangle$ is not a better configuration as far as
in reproducing the lowest energy.

To prove Eq.~(\ref{Equiv}), one can establish by direct calculation
that the color-color part in Eq.~(\ref{eff}) has the same expectation value,
\begin{eqnarray}
\langle \text{C}_i|\sum_{j<k} \lambda_j\cdot \lambda_k |\text{C}_i\rangle =-16\label{color_factor}\ ,
\end{eqnarray}
regardless of the color configurations, $|\text{C}_i\rangle$.
The same result can be seen from Eq.~(6) of Ref.~\cite{Shi:2019dpo}
where the calculation has been performed with the configuration $|\text{C}_2\rangle$ only.
Since $\sum_{j<k}  \lambda_j\cdot \lambda_k$ is simply related to $V_{eff}$ as shown in Eq.~(\ref{eff}),
Eq.~(\ref{color_factor}) indeed proves $V_1=V_2=\cdot\cdot\cdot=V_5$.
The other relation $\langle \Psi| V_{eff}|\Psi\rangle = V_i$ must follow immediately.

Another aspect that can be seen from Eq.~(\ref{Equiv}) is that a stable diquark configuration does not necessarily lead
to an optimal hexaquark configuration with the lowest energy.
The $|\text{C}_1\rangle$, as is shown in Eq.~(\ref{c1}), is the hexaquark configuration constructed from the stable diquark $\bm{6}_c$ only
while the $|\text{C}_2\rangle$ in Eq.~(\ref{c2}) is the configuration built from the less stable diquark $\bar{\bm{3}}_c$.
One might naively expect from a diquark model that $V_1$ is the lowest and $V_2$ is the highest among $V_i$.
The others, $V_3,V_4,V_5$, are expected to lie between the two as they contain the two diquark types as their constituents.
But, Eq.~(\ref{Equiv}) shows that
the $|\text{C}_1\rangle$ configuration is not guaranteed to be the lowest energy state in the hexaquark system.
This is certainly different from the naive expectation from a diquark model.
The main reason behind this is that $V_{eff}$ acts not only on the diquarks
but also on other quark pairs that can be formed from the six quarks.
Similar situation occurs in the tetraquark system~\cite{Kim:2016dfq,Kim:2017yur,Kim:2017yvd,Kim:2018zob,Lee:2019bwi}
where the tetraquark with the spin-1 diquark configuration
turns out to be more stable than the one with the spin-0 configuration
even though the spin-0 diquark is more compact than the spin-1 diquark.

In fact, the result in Eq.~(\ref{Equiv}) is not accidental.
As we have discussed briefly in the last paragraph of Sec.~\ref{subsec:color}, it is a natural consequence
coming from a freedom in dividing
the six quarks into three diquarks in constructing the hexaquark system.
Since $V_{eff}$ acts on all the pairs among the six quarks,
its expectation value must be the same regardless of how the six quarks are divided into three diquarks.

To put it more explicitly, let us rewrite
$|\text{C}_1\rangle$ in Eq.~(\ref{c1}), which was written in the (12)(34)(56) division,
in terms of the new division (13)(24)(56) by moving $q_2,q_3$ in Eq.~(\ref{c1}). We find that
\begin{eqnarray}
|\text{C}_1\rangle &&\Rightarrow \frac{1}{24}\epsilon^{abc}\epsilon^{a'b'c'}(S_{13})_{aa'}(S_{24})_{bb'}(S_{56})_{cc'}\nonumber \\
&&~~-\frac{1}{16}(T_{13})^a(T_{24})^b(S_{56})_{ab}\label{c1a}\ .
\end{eqnarray}
Comparing the RHS with Eq.~(\ref{c1}), (\ref{c5}),
we see that the first term is basically the $|\text{C}_1\rangle$ type in this new division,
which we denote as $|\text{C}^\prime_1\rangle$,
and the second term is the $|\text{C}_5\rangle$ type, which we denote as $|\text{C}^\prime_5\rangle$.
In other words, Eq.~(\ref{c1a}) can be expressed as
\begin{eqnarray}
|\text{C}_1\rangle = \frac{1}{2}|\text{C}^\prime_1\rangle -\frac{\sqrt{3}}{2}|\text{C}^\prime_5\rangle\label{ccc}\ .
\end{eqnarray}
Using this configuration in evaluating $\langle V_{eff}\rangle$,
we find that
\begin{eqnarray}
V_1 = \frac{1}{4} V^\prime_1 +\frac{3}{4}V^\prime_5\label{pv1}\
\end{eqnarray}
where $V^\prime_1,V^\prime_5$ are the expectation values of $V_{eff}$ with respect
to $|\text{C}^\prime_1\rangle, |\text{C}^\prime_5\rangle$, respectively~\footnote{The mixing terms like $\langle \text{C}^\prime_1 | V_{eff} |\text{C}^\prime_5 \rangle$
are found to be zero.}.
But, since $|\text{C}_1\rangle, |\text{C}^\prime_1\rangle$ differ only by the labeling either in (12)(34)(56) or in (13)(24)(56),
both must have the same expectation value for $V_{eff}$. The same thing applies also to $|\text{C}_5\rangle, |\text{C}^\prime_5\rangle$.
As a result, we must have
$V_1 =  V^\prime_1, \quad  V_5=  V^\prime_5$ .
We emphasize that this is a consequence from the arbitrariness in labeling our hexaquark system.
Utilizing this in Eq.~(\ref{pv1}), we arrive at
\begin{equation}
V_1=V_5
\end{equation}
as anticipated from Eq.~(\ref{Equiv}).
Other relations in Eq.~(\ref{Equiv}) can be derived also by taking similar steps.
Note that the equation like Eq.~(\ref{Equiv}) may not be satisfied if the strange quarks are involved in the hexaquark system.

\section{Hexaquark Mass}
\label{sec:mass}

We now present and discuss the hexaquark mass calculated from our wave function.
Plugging Eq.~(\ref{color_factor}) into Eq.~(\ref{eff}), we obtain the formula for $\langle V_{eff}\rangle$ as
\begin{eqnarray}
\langle V_{eff}\rangle = -16\left [ \frac{v_0}{4m^2} + \frac{v_1}{m^2} \right ]  + v_2\ .
\label{expectation}
\end{eqnarray}
Using the constituent quark mass as $m=330$ MeV and the effective parameters ($v_0,v_1,v_2$)
determined from the baryon spectroscopy in Eq.~(\ref{parameters}), we find that
\begin{equation}
\langle V_{eff}\rangle = 361.5~\text{MeV}\label{c-eff}\ .
\end{equation}
This value is positive
suggesting that this hexaquark is a resonance state like the $\Delta$ baryon whose effective potential is also positive. (see Table~\ref{baryon masses}.)

By putting Eq.~(\ref{c-eff}) in the mass formula of Eq.~(\ref{mass}), we finally arrive at our prediction for the hexaquark mass,
\begin{eqnarray}
M_H=2341.5~\text{MeV}\label{c-mass}\ .
\end{eqnarray}
This is indeed very close to the experimental mass of $d^*(2380)$, only 40 MeV below.
Hence, as far as the mass is concerned, the hexaquark picture may not be ruled out from a possible structure for $d^*(2380)$.
In addition, as explained in the previous section,
this mass is the same whether it is calculated with the fully antisymmetric wave function $|\Psi\rangle$, Eq.~(\ref{full}),
or with any of the five configurations $|\psi_i\rangle$, Eq.~(\ref{CI}).
Because of this, Shi et al.~\cite{Shi:2019dpo} could have gotten the same hexaquark mass
if they took our effective potential, Eq.~(\ref{eff}), as well as the mass formula, Eq.~(\ref{mass}), in their calculation
that facilitates $|\psi_2\rangle$ only.
Therefore, in our approach, finding the fully antisymmetric wave function, even though it is needed as a physical state,
is not so crucial in determining the hexaquark mass.

Our result is very different from Ref.~\cite{Park:2015nha} which performed a variational
calculation that takes into account the spatial dependence of the effective
potentials, and obtained
a much larger mass 2630 MeV or 2809 MeV depending on the interaction type adopted in the calculation.
The calculated mass there is too large to be a $d^*(2380)$ mass and thus excludes the hexaquark possibility for $d^*(2380)$.
On the other hand, in our approach, we rely on a simplified effective potential of the contact type, $V_{eff}$ in Eq.~(\ref{eff}), and,
as a result, the five configurations $|\psi_i\rangle$
have the same expectation value as the full wave function does.
No variation is necessary in our simplified approach as all the six quarks are
assumed to be in one spatial point in an average sense.

Nevertheless, it is interesting to compare our result with the mass of the two-baryon state
calculated long ago by Dyson and Xuong~\cite{Dyson:1964xwa}.
Purely from the SU(6) classification, they predicted that the two-baryon mass
in the $(I,J)=(0,3)$ channel is around 2350 MeV,
which is also very close to the $d^*(2380)$ mass.
This old prediction therefore can be used to advocate a different picture for $d^*(2380)$,
the molecular-type resonance composed of two baryons.
In this calculation, there are only two inputs, one is
the deuteron mass and the other is a parameter related to the mass formula for the baryon multiplet.
Their approach, similarly to ours in sprit, relies also on a simplified picture without explicit
spatial dependence of the potential and so on.
Their mass is only 10 MeV higher than our mass in Eq.~(\ref{c-mass}) calculated based on the hexaquark picture with constituent quarks.
Therefore, as far as the mass is concerned,
the two pictures can give a reasonable description for the $d^*(2380)$
although they seem conflicting each other for its internal structure.
Maybe one possible way of reconciling the two pictures can be sought from the fact that the hexaquark picture is more extensive.
In other words, the hexaquark picture can accommodate the two-baryon picture because it includes
the two-baryon state as its component in addition to the hidden-color component.
The resulting similar mass can be understood in our terminology
as having similar value for $\langle V_{eff} \rangle$
whether it is calculated with the two-baryon component or with the hidden-color component.
Deducing from Eq.~(\ref{Equiv}), and also as advocated in Sec.~\ref{sec:veff},  it is certainly possible that the different
configurations have the similar value for $\langle V_{eff} \rangle$.

\section{Summary}
\label{sec:summary}

In summary, we have constructed in this work the hexaquark wave function that might be relevant for the $d^*(2380)$
with the quantum numbers of $I(J^P)=0(3^+)$.
Our assumption in this construction is that
the hexaquark is composed only of $u,d$ quarks that are all in an $S$-wave state.
Since the spatial and spin parts of the wave function are symmetric under any quark exchange,
the rest color-isospin part must be antisymmetric.
The color-isospin part of the hexaquark wave function is constructed first in a diquark
basis where constituting diquarks are being prepared to be antisymmetric in color-isospin.
Note that the diquarks in our work have been adopted as a convenient tool for describing the hexaquarks
so they are not necessarily restricted to the compact types as in the normal diquark models.
There are five color-isospin configurations in this construction. We then construct
a fully antisymmetric color-isospin wave function by linearly combining the five configurations.
It turns out that all the five configurations are equally important to the total wave function.
In particular, all the five configurations
were found to give the same hexaquark mass as the total wave function does.
For the effective potential, we have used the contact type composed of the color-spin, color-electric and
constant shift,
the same type that has been used in the previous works on tetraquarks~\cite{Kim:2016dfq,Kim:2017yur,Kim:2017yvd,Kim:2018zob}.
The empirical parameters associated with the potential are determined from the baryon spectroscopy.
Using these parameters, we have calculated the hexaquark mass to be around 2340 MeV which is quite
close to the $d^*(2380)$ mass, only 40 MeV below.
Therefore, we conclude that the hexaquark picture is still promising as a possible structure for $d^*(2380)$.

In closing, we want to make two remarks.
First is the advantageous aspect of using the diquark basis in comparison with other basis like $(3q)(3q)$ partition.
In particular, it provides a convenient way to construct the hexaquark wave function
that are totally antisymmetric. Namely, the color-part of wave functions are conveniently classified
according to Eqs.~(\ref{c1})$\cdot\cdot\cdot$~(\ref{c5}) and these can be easily incorporated to the isospin parts of
Eqs.~(\ref{I1})$\cdot\cdot\cdot$~(\ref{I5}). This then straightforwardly leads to an explicit expression for the
fully antisymmetric wave function as given in Eq.(\ref{explicit}).
Another thing to mention is the fact that the diquark approach~\cite{Shi:2019dpo} whose wave function is not fully antisymmetric
still yields the same hexaquark mass that can be obtained from the fully antisymmetric wave function.
As discussed in Sec.~\ref{sec:veff}, this result is originated from a freedom in dividing the six-quark into three diquarks
in constructing the hexaquark system.

\acknowledgments

\newblock
The work of H. Kim and K.S.Kim was supported by the National Research Foundation of Korea(NRF) grant funded by the
Korea government(MSIT) (No. NRF-2018R1A2B6002432, No. NRF-2018R1A5A1025563).
The work of M. Oka is supported by a JSPS KAKENHI Grants No. JP19H05159.

\appendix*
\section{Determination of $v_0, v_1, v_2$ from baryon spectroscopy}
\label{sec:baryon}

The effective potential, $V_{eff}$ in Eq.~(\ref{interaction}), contains three parameters, $v_0, v_1,v_2$,
which are related to the color-spin, color-electric potential, and the constant shift.
In this appendix,
we determine these parameters from the mass formula Eq.~(\ref{mass}) when it is applied to the baryon octet and decuplet.
The inputs in this analysis are the constituent quark masses which we set as $m_u=m_d=330$ MeV, $m_s=500$ MeV.


\begin{table*}[t]
\centering
\begin{tabular}{c|c|c|c|c||c|c|c|c}  \hline\hline
\multirow{2}{*}{Baryon}  & \multirow{2}{*}{$M_\text{expt}$} &\multicolumn{3}{c||}{$M_\text{theory}$} &\multicolumn{4}{c}{Each term in Case III}\\
                               \cline{3-9}
                 &                &  Case I    &  Case II &  Case III     & $\sum m_q$ & $V_{CS}$ & $V_{CE}$ & $v_2$          \\
\hline
$N$              & 940                 & 844    & 940 (input)  & 940 (input) & 990 & $-146.0$& $-26.5$ &  \\
$\Delta$         & 1232                & 1136   & 1232 (input) & 1232 (input) & 990 & 146.0& $-26.5$ & \\ 
$\Lambda$        & 1116                & 1014   & 1088         & 1116 (input) & 1160 & $-146.0$& $-20.5$ & \\
$\Sigma$         & 1193                & 1080   & 1154         & 1182         & 1160 & $-79.8$& $-20.5$ & 122.5\\
$\Sigma^* $      & 1385                & 1273   & 1279         & 1375         & 1160 & 112.9& $-20.5$ & \\
$\Xi$            & 1320                & 1223   & 1347         & 1330         & 1330 & $-107.3$& $-15.5$ & \\
$\Xi^*$          & 1531                & 1415   & 1472         & 1522         & 1330 & 85.4& $-15.5$ & \\
$\Omega$         & 1672                & 1564   & 1605         & 1675         & 1500 & 63.6& $-11.5$ & \\ \hline
\end{tabular}
\caption{Calculated baryon masses in three different cases as well as experimental masses are reported here.
See the text for details in calculating the masses in each case.
The last four columns show individual contribution to the baryon mass from the quark mass,
color-spin, color-electric, and the constant shift in Case III.  The constant shift is of course the same in all the channels.
All the numbers are given in MeV unit.
}
\label{baryon masses}
\end{table*}

Using the effective potential Eq.~(\ref{interaction}) and the appropriate wave functions constructed
for the baryon octet and decuplet,
it is straightforward to derive the mass formulas,
\begin{eqnarray}
m_N &=& 3m_u+2\frac{v_0}{m^2_u}-8\frac{v_1}{m^2_u}+v_2\label{nucleon}\ ,\\
m_{\Delta} &=& 3m_u-2\frac{v_0}{m^2_u}-8\frac{v_1}{m^2_u}+v_2\label{Delta}\ ,\\
m_{\Lambda} &=& 2m_u + m_s + 2\frac{v_0}{m^2_u}\nonumber\\
&&
-\frac{8}{3}v_1\left[\frac{1}{m^2_u}+\frac{2}{m_u m_s}\right] +v_2\label{Lambda}\ ,\\
m_{\Sigma} &=& 2m_u + m_s - \frac{8}{3} v_0\left [\frac{1}{4m^2_u}-\frac{1}{m_um_s}\right ]\nonumber\\
&&-\frac{8}{3}v_1\left[\frac{1}{m^2_u}+\frac{2}{m_u m_s}\right] +v_2\label{Sigma}\ ,\\
m_{\Sigma^*} &=& 2m_u + m_s - \frac{8}{3} v_0\left [\frac{1}{4m^2_u}+\frac{1}{2m_um_s}\right ]\nonumber\\
&&-\frac{8}{3}v_1\left[\frac{1}{m^2_u}+\frac{2}{m_u m_s}\right] +v_2\label{Sigma*}\ ,\\
m_{\Xi} &=& m_u + 2m_s - \frac{8}{3} v_0\left [-\frac{1}{m_um_s}+\frac{1}{4m_s^2}\right ]\nonumber\\
&&-\frac{8}{3}v_1\left[\frac{2}{m_um_s}+\frac{1}{m_s^2}\right] +v_2\label{Xi}\ ,\\
m_{\Xi^*} &=& m_u + 2m_s - \frac{8}{3} v_0\left [\frac{1}{2m_um_s}+\frac{1}{4m_s^2}\right ]\nonumber\\
&&-\frac{8}{3}v_1\left[\frac{2}{m_um_s}+\frac{1}{m_s^2}\right] +v_2\label{Xi*}\nonumber\\
m_{\Omega} &=& 3m_s-2\frac{v_0}{m^2_s}-8\frac{v_1}{m^2_s}+v_2\label{Omega}\ .
\end{eqnarray}

The three parameters $v_0$, $v_1$, $v_2$ appearing in $V_{eff}$ will be fixed in three different scenarios.
In the first scenario (Case I), we consider the color-spin potential only
by setting $v_0\ne 0, v_1=v_2=0$ in Eq.~(\ref{interaction}).
It is well known that the color-spin potential can explain the
mass differences between the baryons with
the same flavor content, namely $\Delta M_{exp}\approx \Delta \langle V_{CS}\rangle$ (see Table~IV of Ref.~\cite{Kim:2014ywa}).
Only input in this case is the mass difference of $N,\Delta$. Taking the difference between Eq.~(\ref{nucleon}) and Eq.(\ref{Delta}),
we find $m_{\Delta}-m_N =-4 v_0/m^2_u$ which fixes
\begin{equation}
v_0=(-199.6~\text{MeV})^3\label{parameter1} .
\end{equation}
if we use the experimental masses $m_\Delta, m_N$ as inputs.
The baryon masses determined in this scenario are given in the third column in Table~\ref{baryon masses}.
The calculated masses are about 100 MeV less than the experimental masses.
So the color-spin interaction alone, even though it is successful in reproducing the mass differences,
is not precise enough to generate the experimental masses.

In the second scenario (Case II), we include the color-electric potential in addition to the color-spin potential
($v_0\ne 0, v_1 \ne 0, v_2 = 0$).
The additional parameter $v_1$ is fixed from Eq.~(\ref{nucleon}) by using $m_N=940$ MeV as an input.
That is,
\begin{eqnarray}
&&-8\frac{v_1}{m^2_u}=m_N-3m_u-2\frac{v_0}{m^2_u}=96~\text{MeV}\label{alpha}\nonumber \\
&&\rightarrow v_1=(-109.3~\text{MeV})^3\label{v1_1}\ .
\end{eqnarray}
The baryon masses determined in this case are given in the 4th column in Table~\ref{baryon masses}.
This result is much better than Case I although the calculated masses deviate
from the experimental masses maximum up to 100 MeV.

In the third scenario (Case III), we include all the three potentials, color-spin, color-electric and constant shift,
($v_0\ne 0, v_1 \ne 0, v_2\ne 0$). In this case, we use $m_\Lambda=1116$ MeV as an additional input.
Plugging the input values of $v_0, m_N, m_\Lambda$ in Eqs.~(\ref{nucleon}), (\ref{Lambda}),
we find the two constraints,
\begin{eqnarray}
&&-8\frac{v_1}{m^2_u}+v_2=96~\text{MeV}\ ,\\
&&-\frac{8}{3}v_1\left(\frac{1}{m^2_u}+\frac{2}{m_u m_s}\right) +v_2=102~\text{MeV}\ ,
\end{eqnarray}
which lead to
\begin{eqnarray}
v_1=(71.2~\text{MeV})^3\ ; \quad v_2=122.5 \text{MeV}\label{parameter2}\ .
\end{eqnarray}
The baryon masses in this scenario, which are given in the 5th column in Table~\ref{baryon masses},
have an excellent agreement with $M_\text{expt}$ within 10 MeV.  This also shows that the effective potential, Eq.~(\ref{interaction}),
as well as the empirical parameters $v_0,v_1,v_2$ given in Eqs.~(\ref{parameter1}),(\ref{parameter2}) are
successful in describing the baryon spectroscopy.

We also examine the relative importance of each potential term in Case III
by separating the calculated mass into the quark mass term, color-spin term,
color-electric term, and the constant shift.
They are listed in the last four columns of Table~\ref{baryon masses}.
A common feature is that the quark mass term is the biggest as it should be in the constituent quark picture.
The color-spin potential is the second biggest for  $N,\Delta$,
but its contribution becomes less important in the resonances with strangeness
mostly because the color-spin interaction is proportional to $1/m_i m_j$ as in Eq.~(\ref{interaction}).
Relating to this is the relative contribution of the constant shift, which is slightly less than the color-spin term in $N,\Delta$,
becomes the second biggest in most resonances with strangeness.
But in all resonances, the color-electric term contributes marginally in generating the baryon masses.

Another thing to mention is that the parameters in Eqs.~(\ref{parameter1}),(\ref{parameter2})
are determined from a baryon system of $qqq$.
This system is similar to the hexaquark system in a sense that both are composed of quarks only without antiquarks.
This is in contrast to the tetraquark system which is composed of quarks and antiquarks.
Indeed, the $v_0$ value determined from the tetraquark system is $v_0\approx (-193)^3$ MeV$^3$,
slightly less than Eq.~(\ref{parameter1}).
Even though the difference is rather small, our parameters
determined from the baryon system must have better justification when they are applied to hexaquarks.

\end{document}